# Emotional Interaction Qualities: Vocabulary, Modalities, Actions, And Mapping

Albrecht Kurze

Chemnitz University of Technology, Albrecht.Kurze@informatik.tu-chemnitz.de

Have you ever typed particularly *powerful* on your keyboard, maybe even *harsh*, to write and send a message with some emphasis of your emotional state or message? Did it work? Probably not. It didn't affect how you typed or interacted with your mouse. But what if you had other, connected devices, with other modalities for inputs and outputs? Which would you have chosen, and how would you characterize your interactions with them? We researched with our multisensory and multimodal tool, the *Loaded Dice*, in co-design workshops the design space of IoT usage scenarios: what interaction qualities users want, characterized using an interaction vocabulary, and how they might map them to a selection of sensors and actuators. We discuss based on our experience some thoughts of such a mapping.

## 1 INTRODUCTION

Some years ago we designed and developed the *Loaded Dice* [9,10], a multisensory and multimodal hybrid toolkit to ideate IoT devices and scenarios, e.g. for the 'smart' home, and with different groups of co-designers [3,8,9]. The *Loaded Dice* filled a gap between analog, non-functional tools, often card-based, e.g. *KnowCards* [1], and functional but tinkering based tools, e.g. *littleBits* [2], for multisensory and multimodal exploration, ideation and prototyping.

We will introduce a) our adapted and extended interaction vocabulary, b) how we use it in a method to explore and the describe the WHY and HOW of interactions with and through connected devices; and c) we introduce the Loaded Dice, what sensing and actuating functions they have, and how we use the to let participants map interaction qualities to modalities to align the WHY and the HOW of interactions.

This brings us to our core question: *Is it possible to derive a (somehow universal) mapping between certain interaction qualities, i.e. emotional ones, and specific modalities and actions?*

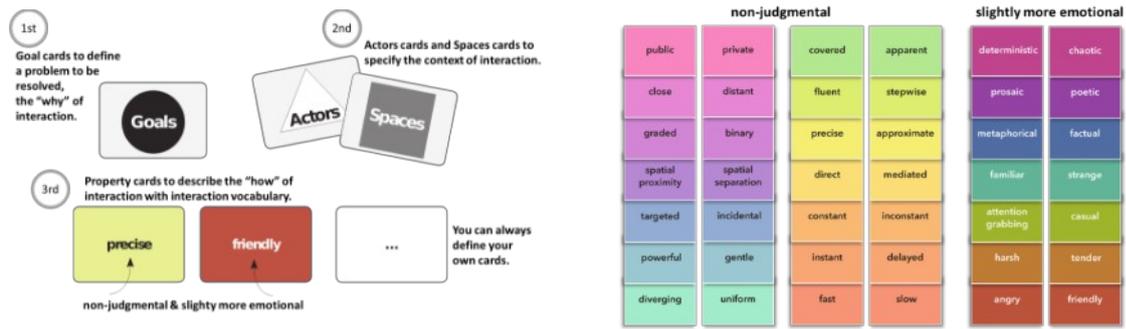

Figure 1: left: the types of cards of our co-design method, 1st setting a goal (why), 2nd context of interaction (actors and space), 3rd defining desired interaction qualities; right: cards with terms of the extended interaction vocabulary

## 2 ADAPTED AND EXTENDED INTERACTION VOCABULARY

Diefenbach [4] introduced in 2013 a first interaction vocabulary to describe interaction qualities in a user perspective. The original vocabulary consisted of 11 pairs of adjectives and antonyms, e.g. *fast* and *slow*. An example of how the vocabulary was intended to describe interaction qualities: When we switch the light in a room, this happens in a *binary* way at the switch (on/off) with an *instant* effect in the same way at the lamp *distant* to the switch. With a dimmer the input and output are *graded* in a *fluent* or *stepwise* manner (vocabulary terms in italics). The original vocabulary was intended for use as a semantic differential on a graded scale, e.g. in a questionnaire.

Our intention was to not only characterize interactions with a single object but also in complex connected interaction scenarios. In scenarios as the IoT allows them for smart connected things, across multiple devices and shared between multiple involved actors (typically human users but not limited to them).

While Diefenbach's intention for the original interaction vocabulary was first to describe "the HOW of interaction" [4] they also had drawn a first conclusion between HOW and WHY of interaction. We put this first. It became clear to us that it is often not meaningful to isolate the HOW from the WHY of interaction. Therefore we embedded the interaction vocabulary methodically in a goal-actors-properties driven scenario creation to match the IoT design space [3]. We adapted and extended the original vocabulary. We did this in the same way as the original vocabulary was constructed – as pairs of adjectives and antonyms. Additionally, we introduced to the vocabulary an extension with some more emotional terms to grasp interaction qualities often beyond a non-judgmental dimension [4]. We discuss their role following on with mappings to specific modalities. We have already iterated the actual terms based on what we have learned in co-design workshops using the vocabulary. We see the vocabulary still as work in progress.

We created based on the vocabulary a set of cards for the use in co-design workshops. On the front face an adjective and on the back the antonym (fig. 1b). We introduced a subtle differentiation between the two categories. The non-judgmental terms (including the original terms) are in black letters on colored background while the slightly more emotional terms are in white letters. In contrast to Diefenbach's graded semantic differential we decided with the cards only for the extremes - an 'either or'. However, this stimulates in the co-design workshops a verbalization how something is meant – often not in the extremes but then user defined graded.

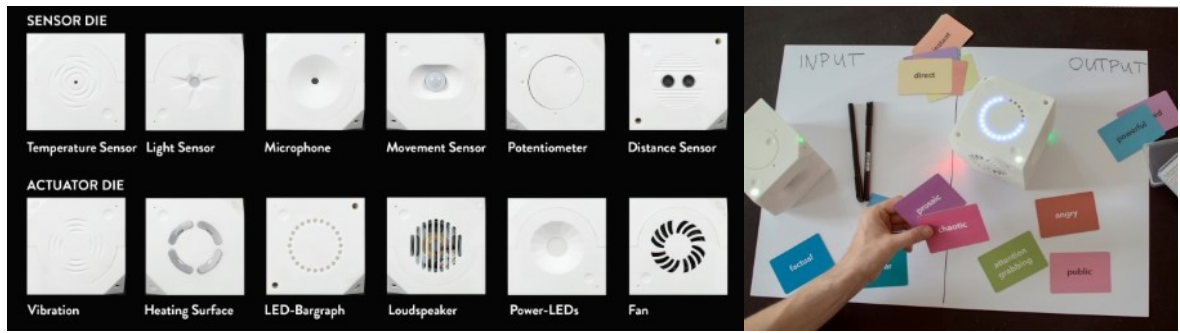

Figure 2: left: faces and functions of the Loaded Dice – sensors and actuators [9]; right: an example of using the cards to characterize and map input and output qualities of an ideated connected product with help of the Loaded Dice [6]

## 3 INTERACTION MODALITIES

The *Loaded Dice* are a set of two cubical devices wirelessly connected (fig. 2a). Each cube has six sides, offering in one cube six sensors and in the other cube six actuators, one on each side, suitable for multisensory and multimodal environmental and user interactions. The sensor cube normalizes a raw sensor value meaningfully, transmits it, and then the other cube actuates it mapped on an output. The cubical shape communicates the intuitive reading that the top side is active, like a die, offering an easy and spontaneous way to re-combine sensors and actuators. Every sensor-in and actuator-out combination is possible resulting in 36 combinations in total. [6] New multisensory interaction modalities, not yet implemented, e.g. smell, have the potential to broaden interaction qualities even further and especially in an emotional way [7].

## 4 MAPPING INTERACTION QUALITIES

Last step in our co-design method is a mapping of desired interaction qualities by participants to interaction modalities represented by the *Loaded Dice* (fig. 2b). The *extended interaction vocabulary* allows characterizing the intended interactions very well while the *Loaded Dice* allow participants to try them out up to a certain degree. This way our workshops often brought up a number of unconventional ideas of multisensory interactions with devices, often far beyond ordinary inputs and especially outputs. In our experience sensory sensations and modalities do not need to be perfect, at least for ideation of interactions and scenarios as well as for mapping interaction qualities. It is about bringing the idea and the core concept behind it to the co-design activities. A demonstration of a technical possibility for sensing and actuating as a stimulus is often enough to trigger further thinking and verbalization of how something might be used.

We found some repeating themes when it comes to a mapping between certain interaction characteristics and suitable sensing as well as actuating possibilities. For example, the thermo-element was not only associated with *slow* and warmth literally but also with 'love' and *tender* in a *poetic* way, while loud sound and bright light were selected for *powerful*, *attention-grabbing* and sometimes even *harsh* interactions etc. Participants often chose non-visible and non-audible modalities for *private* interactions, *covered* and not easily perceivable by others, only noticeable to a mentioned one, e.g. using heat or vibration in ideated wearable devices. In another case participants mapped the vibration motors and the associated sound caused when having the Loaded Dice placed on a wooden table to *attention-grabbing* and *harsh,* associated to feelings of being alarmed and named it "electronic rattlesnake".

We also found similar patterns for inputs. The distance sensor can detect a hand in proximity in different ways. In a *graded* kind, if done *slowly*, allowing for *gentle* gestures, e.g. swiping with the hand through the air above the sensor, without touching something, without any force. As these gestures can be very similar to petting something they were associated with this action in a *poetic* and very *tender* way. On the other hand, a fast and sudden movement is also detectable, like a punch, being very *powerful*, *targeted* and *harsh*. While the distance sensor allows for such a differentiation based on the speed of hand movement the PIR movement detection sensor allows not – what also might be wanted, e.g. for an only *binary* type of input.

Both ways of using the distance sensor for proximity-based hand gestures are possible and meaningful. However, the HOW of the interaction is then depending on the emotional state of the user and the WHY of interaction. Therefore, it makes a difference what a user tries to express and in an end-to-end view of interactions 'through' devices [6], from one device to another device, as communication to another actor. Has the user the intention to send a message with a positive emotion, a non-verbal equivalent of "I love you tender", or to send a message associated with a negative emotion, e.g. with an equivalent in "Turn the damn music down"? In terms of Hassenzahl's model of interactions for experience design [5], as a hierarchy of WHY, WHAT, and HOW of interactions, it is clear that this WHY, the motive, defines the WHAT and the HOW. Therefore, it will not be a simple one-dimensional mapping between an emotional interaction quality and specific sensor / actuator modalities. The motive of use (as higher-level use goal) and the expressed or implied associate actions (e.g. petting or punching) must also be considered.

## 5 CONCLUSION

While we see especially big potential in the use of an interaction vocabulary and different modalities for intending or expressing emotional interaction qualities, it still needs further exploration to identify certain patterns for a mapping.

## ACKNOWLEDGMENTS

This research is funded by the German Ministry of Education and Research (BMBF), grant FKZ 16SV7116.